# The activity of native vacuolar proton-ATPase in an oscillating electric field – demystifying an apparent effect of music on a biomolecule

Pál Petrovszki, Krisztina Sebők-Nagy, Tibor Páli*

Institute of Biophysics, Biological Research Centre, Eötvös Loránd Research Network, Temesvári krt. 62, 6726 Szeged, Hungary

*Email: tpali@brc.hu .

**Abstract**: The effect of an oscillating electric field generated from music on yeast vacuolar proton-ATPase (V-ATPase) activity in its native environment is reported. An oscillating electric field is generated by electrodes that are immersed into a dispersion of yeast vacuolar membrane vesicles natively hosting a high concentration of active V-ATPase. The substantial difference in the ATP hydrolysing activity of V-ATPase under the most stimulating and inhibiting music is unprecedented. Since the topic, i.e. an effect of music on biomolecules, is very attractive for non-scientific, esoteric mystification, we provide a rational explanation for the observed new phenomenon. Good correlation is found between changes in the specific activity of the enzyme and the combined intensity of certain frequency bands of the Fourier spectra of the music clips. Most prominent identified frequencies are harmonically related to each other and to the estimated rotation rate of the enzyme. These results lead to the conclusion that the oscillating electric field interferes with periodic trans-membrane charge motions in the working enzyme.

**One-Sentence Summary**: The enzymatic activity of native vacuolar proton-ATPase in an oscillating electric field generated from music depends on the presence of frequency bands in the Fourier spectra of the music clips that are harmonically related to the estimated rotation rate of the enzyme.

**Introduction**

Music is made by humans for humans and beyond the obvious emotional effects, the psychological, physiological, and neurobiological effects of music on humans are well studied[1–4]. There is also a long-standing interest in the effect of music on non-conscious living matter. In most studies on organisms (without any hearing organ), or biomolecules and life processes they are subjected to "listening" music from the air, and the observed effects are rather small and indirect[5–9]. The first objective of this study is to test a biochemical effect, if any, of a time-dependent physical quantity (other than pressure) derived from music. We follow a unique approach: we convert music to alternating current and that to oscillating electric field (shortly AC field) and measure how the activity of an enzyme changes under the effect of that field. Yeast vacuolar proton-pumping adenosine-triphosphate (ATP) hydrolase (vacuolar proton-ATPase or V-ATPase) is an optimal choice because it plays crucial roles in many life processes[10–13] and it exhibits ATP hydrolysing and trans-membrane proton pumping activity coupled by a rotary mechanism that involves periodic charge movements[11,13–16]. It has been known for decades that the activity of enzymes embedded in native or model membranes can be altered even by weak AC field because the field is amplified by the membrane, provided that the cells or vesicles are not leaky. These studies have hitherto utilised a regular, most commonly sine waveform[18–21]. We have also shown recently that V-ATPase is sensitive to AC field of pure sine waveform[16]. However, pure sine oscillation hardly ever happens in living cells or in nature in general, and it has been suggested that the pure sine may not be the most efficient waveform in such studies[17]. Therefore, as our second objective, we address the question, for the first time: Can an AC field with a complex, non-regular time-dependence and a wide and variable frequency spectrum affect enzymatic activity? Apart from the interest in the effect of music on living matter, we have chosen music as the source of the AC signal because, although it is also not present in living cells (except perhaps the those related to hearing) it satisfies these requirements very well.

V-ATPase belongs to the family of membrane-attached biological macromolecules whose functioning involves true, full cycle rotation of some parts, called rotor, relative to other parts, called stator[10-13,22–25]. The normal function of V-ATPase is to pump protons across certain biomembranes and it is a key rotary enzyme in all eukaryotic cells, acidifying intracellular compartments and the extracellular space in some tissues[13,23,25,26]. A class of proteins including the proteolipid c-ring sub-complex (the main part of the rotor) has functions independent from V-ATPase[27–30]. V-ATPase is also a potential therapeutic target in several diseases[13,26,31,32]. V-ATPase works in the opposite sense as the better known F-ATP synthase[22–24,33–39]. Proton transport by V-ATPase is energised from the chemical energy stored in ATP via binding and hydrolysis, which is converted into mechanical force rotating a group of certain subunits relative to the rest of the protein complex. F- and V-ATPases are true molecular engines[23,33,35,37,40]. In both enzymes, ATP hydrolysis (or synthesis) and proton transport are strongly coupled via the rotary mechanism[13,24,25, 31,32,35,36]. Since the periodicity of vectorial charge movements is related to that of the rotation, an oscillating electric field (EF) with the frequency of the turnover of the main charge movements should have a maximum effect on enzymatic activity.

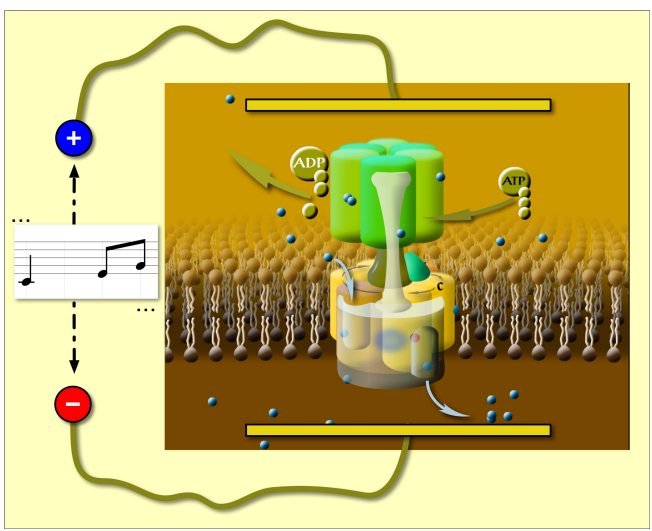

**Figure 1.** The oscillating trans-membrane electric field is parallel to the direction of the net vectorial proton transfer in V-ATPase (based on refs. 15 and 16).

Indeed, in a first of its class measurement[16] we applied AC field on a rotary enzyme (the V-ATPase) and discovered a new resonance-like frequency response of the enzyme to AC field, which allowed us to directly determine the rotation rate in native V-ATPase in its native membrane environment, without any genetic or structural modifications. In that experimental setup, most water soluble components in the vacuoles were removed by washing, yielding membrane vesicles with high concentration of native V-ATPase. Two flat platinum electrodes were immersed in the vesicle suspension, in a flat cuvette. If the vesicles are not leaky the EF sensed by membrane proteins (Fig. 1) is amplified by $1.5 * \cos(\theta) * R_0/d$ times relative to that in the aqueous phase ($R_0$ and $d$ are the radius of the vesicle and the thickness of the lipid bilayer, respectively, and $\theta$ is the angle between a line normal of the membrane surface with the EF, at the point of interest)[18]. It has also been shown by Astumian and co-workers that an AC field in the order of 20 V/cm amplified by tightly sealed membrane vesicles was strong enough to increase the activity of membrane-bound enzymes and even induce active transport by ATPases[18-20]. We have shown recently that making the vesicles leaky with pore forming ionophores removes the effect of the AC field[16] on V-ATPase (but not its control activity) in our vacuolar vesicle preparations, proving that our vesicles are sealed and the membrane-amplified AC field acts on V-ATPase-related charge movements along the membrane normal. In the present study, the same experimental setup was used but with music in place of a sinusoidal waveform. Concanamycin A (ConcA) was used to determine the V-ATPase contribution to the total ATPase activity because it is a very specific and highly potent inhibitor of this particular enzyme. It binds to the intramembranous domain of V-ATPase and blocks rotation, hence proton transport and ATP hydrolysis[31,32,41].

**Table 1**. The content of the audio clips* played to the V-ATPase as oscillating electric field, sorted from maximum enhancement (top) to maximum reduction (bottom) of the specific activity of the enzyme.

| Audio clip | Content |
| --- | --- |
| 0_jarret | Jazz, piano: K. Jarret: The Köln concert, Part IIa (*D major*) |
| 1_deszk | Folk songs for wine drinking, from West Hungary (*D major, A minor, A major; the pitch was not stable during the performance*) |
| 2_mozart | W.A. Mozart: Divertimento D-dur, KV251, Rondeau, Allegro Assai (*D major*) |
| 3_bach | J.S. Bach: Brandenburg Concertos, 5 in D-major, Allegro (*D major*) |
| 4_deszk | Bagpipe-related folk songs from Nord Hungary (*A major >> A minor; bagpipe in the key of A is ~20 cent flat*) |
| 5_legedi | Hungarian Csángó folk song: Gergely dance (*G harmonic minor or C minor, solo instrument in key of C is ~20 cent sharp, bass drum plays in C during almost the whole clip*) |
| 6_mozart | W.A. Mozart: Eine kleine Nachtmusik, 1, Allegro (*G major*) |
| 7_berry | C. Berry: Back in the USA (*Eb major, but in 430Hz tuning*) |
| 8_boneym | Boney M: Megamix (*C major, C minor, E minor*) |
| 9_bach | J.S. Bach: Toccata and Fugue in d-minor (*D minor*) |
| 10_fchoir | Female choir: rearranged Hungarian folk songs (*G minor >> D minor > G mixolydian*) |
| 11_liszt | F. Liszt: Buch der Lieder für Piano allein, No. 1 (*E major >> G major > F-sharp major*) |
| 12_liszt | F. Liszt: Concerto for Piano & Orchestra in E flat major, 3, Andante (*E flat major*) |
| 13_wnoise | white noise (not music) |
| 14_abba | ABBA: Waterloo (*D major*) |
| 15_jarre | Jean-Michel Jarre: Arpegiator (synthesizer music) (*C minor*) |
| 16_pnoise | pink noise (not music) |
| 17_bach | J.S. Bach: Brandenburg Concertos, 2, in F-major, Allegro (*F major*) |
| 18_deszk | Wedding folk music from South Hungary (*A minor >> A major, C major*) |
| 19_vivaldi | A. Vivaldi: The Four Seasons, Autumn, 1, Allegro (*F major*) |

*Up to three of the most frequented key(s) and scale(s) are also given for each clip (referencing to 440Hz tuning, unless indicated otherwise). Detailed source data of the audio clips are given in Table S1 of the Supplementary Information (linked to the on-line version of the paper).

## Results and Discussion

Table 1 lists the audio clips that were played to the platinum electrodes immersed in the aqueous suspension of yeast vacuolar vesicles with high concentration of active V-ATPase[15,16]. An exhaustive and representative sampling of all music is obviously not possible, the more so since we could not know what musical characteristics would be important, if any. Therefore the sampling is subjective, but we had the following preferences: cover different dynamics and dominant pitch regions, include several genres and have different examples from some of the albums. Amongst those audio clips, white and pink noise are also included. A detailed description of the source of each audio clip is given in the Supplementary Information (linked to the on-line version of the paper). All audio clips were normalised to the same root-mean-square intensity. Double subtractions, released inorganic phosphate form ATP under AC field (in the absence minus presence of ConcA) minus the same in the absence of AC field, were undertaken for each set of experiments with each audio clip, yielding the change in the specific V-ATPase activity presented in Fig. 2 (grey and black squares denote independent experiments and their means, respectively). The music clips are listed in Table 1 in the order of the effect, that is from enhancement to reduction of V-ATPase activity. All the original experimental *OD* and *SA* data are presented in Table S2 in the Supplementary Information. Taking into account that subtraction increases relative error, the effect of music on V-ATPase is substantial. The mean difference of *OD* in the absence and presence of ConcA, without AC field (the control) was $OD_0$ = 0.6126±0.0042 (s.e.m.). The total protein per sample was set to 0.3 mg (verified with the Lowry protein assay)[15,16] and using our inorganic phosphate calibration on the same vacuolar vesicle system[15] the specific V-ATPase activity is $SA_0$ = ~45.2 nmole min$^{-1}$ mg$^{-1}$ in the absence of any AC field, i.e. no music case, corresponding to the zero line in Fig. 2. It is evident from Fig. 2 that the difference in specific V-ATPase activity between the most stimulating and most inhibiting music is more than ~10 times the experimental error and amounts to ~50% of V-ATPase activity under no AC field conditions.

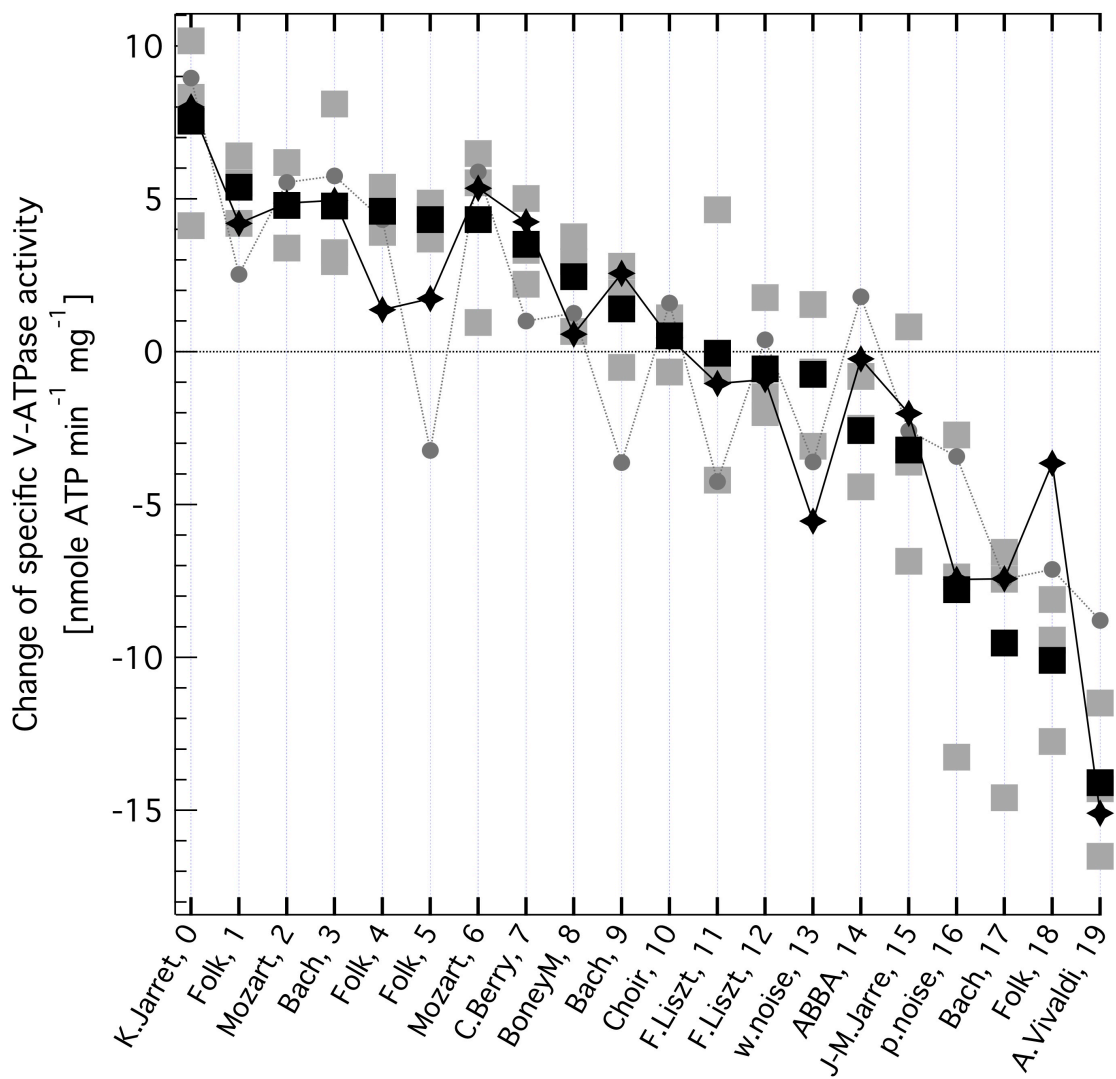

**Figure 2.** Change in the specific enzymatic activity of V-ATPase, as a consequence of exposing the enzyme to oscillating electric field generated from music. Changes are determined relative to the control, no AC field values. The zero line corresponds to the mean control (no AC-field) V-ATPase activity of 45.2 nmole ATP min-1 mg-1. The music sources from various genres and their audio file indexes are indicated on the x axis (see Table S1 in the Supplementary Information for more details). Grey squares are independent experiments, black squares are their mean values. (All optical density (OD) and specific activity (SA) data are given in Table S2 in the Supplementary Information.) The summed intensity of the base frequency at 141.9 Hz and its first 4 overtones (grey circles) and that of 6 selected bands from an iterative search (black diamonds, see text) of the Fast Fourier Transform (FFT) spectra of the normalised music audio clips are also shown after best fitting linear scaling.

As seen in Fig. 2, some of the music clips are stimulating whereas others are inhibiting the enzyme. This is different from results using the sine waveform, which was inhibitory over the whole audio frequency range except for a narrow region[16].

We could not identify any obvious audible characteristics of the music that could account for the preference order of V-ATPase. It can not be related to musical genres or performers, composers since there are cases where clips from the same music CD and performer or composer, e.g., some folk music or J.S. Bach, respectively, landed far from each other in the "preference" list of the enzyme. However, in the music most stimulating V-ATPase, an excerpt from Keith Jarret's famous Köln concert (jazz, piano), a single note dominates during the 5 min clip. That musical note is ~D3 (in 440 Hz tuning) at a frequency of ~147 Hz. Consequently, as shown in Fig. 3 (top), this frequency not only gives the highest peak in the Fast Fourier Transform (FFT) spectrum of this clip but the four next highest peaks are its first four overtones (within the tolerance of the bandwidths). The most inhibitory clip is Vivaldi's Autumn from The Four Seasons. Its spectrum (Fig. 3, bottom) shows a dominant peak at ~173 Hz, which is a frequented musical note in that piece, F3 (in 440 Hz tuning), probably from a cello, and there are no strong overtones present in the spectrum. These observations suggest that the ~173 Hz band is inhibiting and the ~147 Hz band, and possibly some of its overtones, are stimulating V-ATPase, respectively. Our previous calibration[15] can be used to estimate the mean rate of the 60-degree rotation steps of V-ATPase[16] in the absence of AC field, using the above no-AC *OD* value (and assuming that the concentration of the active enzyme is the same as in that study[16]): it is ~59 Hz in the absence of AC field (corresponding to the zero line on Fig. 2) and ~69 Hz under the most stimulating music clip. Interestingly, the above D3 note is reasonably close to the first overtone of this rotation frequency. The above observations prompted us to check the musical keys and scales of the clips, which are indicated in Table 1 too (more details are given in Table 2S of Supplementary Information). A striking observation is that the top four clips stimulating V-ATPase are exclusively or, in one case, in part in the musical key of D, and the next three clips are in the harmonically closest keys of A and G. The bottom end of the list is less consistent in this respect, but the five most inhibitory clips include the pink noise, two clips are in key of F and one clip is in the harmonically closest key (C). There is no clear trend, as concern musical keys and scales, between the top and bottom regions of the list, and it is obvious that these musical qualifiers alone can not explain the direction and magnitude of the AC effect.

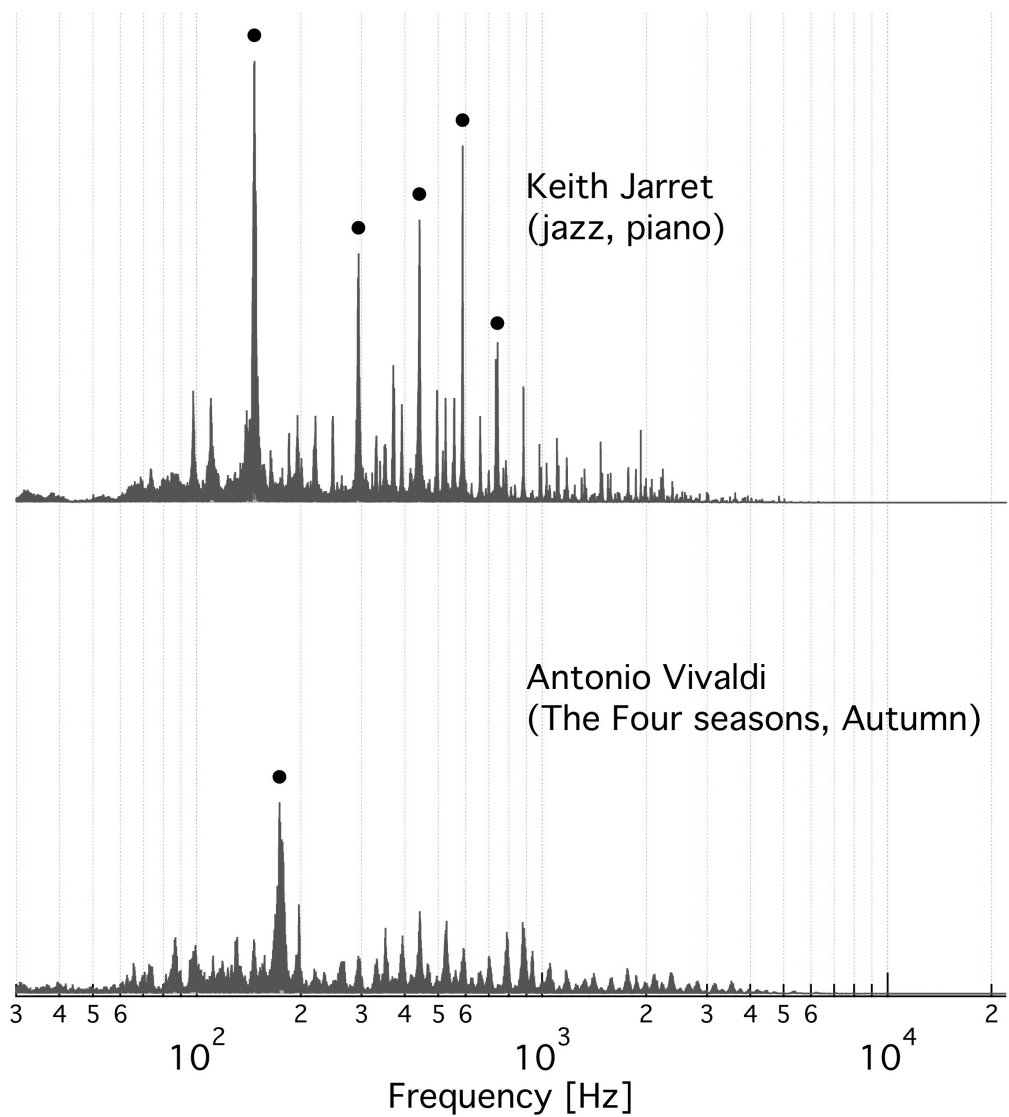

**Figure 3.** The FFT spectrum of the music clip that most stimulates (top) and most inhibits (bottom) the specific activity of V-ATPase. Top: An excerpt from Keith Jarret's Köln concert (jazz, piano). The black circles indicate the five highest peaks at 146.7±1.9 Hz, 293.4±3.3 Hz, 440.9±2.9 Hz, 587.9±1.9 Hz, and 742.5±1.2 Hz. Bottom: An excerpt from the Autumn of Antonio Vivaldi's Four Seasons. The black circle marks the peak at 174.9±5.3 Hz. The ± frequency intervals indicate the line-width of a Gaussian function fitted to the respective peak.

We have done extensive time-domain analysis of the music audio clips (not shown) but that provided no insights vis-à-vis the "preference list" of V-ATPase. Therefore, in order to find more precise explanation for the observed "musical preference" of V-ATPase, we tested to discern whether single frequency bands correlate with the experimental effect. This was accomplished by computing the summed intensity $I(f)$ of a band centred around frequency $f$ for each audio clip, and the series of 20 intensities was linearly fitted via least squares to the experimental data to minimise $\sum(\Delta SA_i - (a + b * I_i(f))^2$, where $i$ indexes the independent experiments, and $a$ and $b$ denote fitting parameters. The linear correlation coefficient between the fitted intensities and experimental data was then calculated. It should be noted that if the best fitting $b$ parameter is negative, the corresponding frequency band correlates negatively with experimental data, i.e., more of that frequency in the spectrum leads to stronger inhibitory effect on the enzyme. Fig. 4 (top) shows the linear correlation coefficient as a function of the frequency of a single frequency band. The effect of the bandwidth was also tested by plotting the correlation coefficient as a function of the bandwidth (not shown) and Fig. 4 (top) was constructed with a fixed bandwidth of 1.4 Hz yielding the highest global maximum. Four of the six prominent positive peaks are harmonically related to (are integer multiples of) a base frequency at 146.7±1.9 Hz, whereas the global maximum at 338.9±4.8 Hz is close to the first overtone of the frequency, 175.4±4.9 Hz at the global minimum. (The ± frequency intervals indicate the line-width of a Gaussian function fitted to the respective peak.) These latter observations suggest that, in addition to the weights, overtone phase is also important. It should also be noted that most of the effective high frequency bands are inhibitory, and all frequencies below 46 Hz and above 11 kHz are strongly and weakly inhibitory, respectively (not shown).

Whether specific harmonic structures (including the first few even, odd, or all overtones) yield a better linear correlation was then tested. Fig. 4 (bottom) shows the linear correlation coefficient between the experimental $\Delta SA_i$ values and the least-squares fitted combined spectral intensity of a base frequency plus its first three overtones as a function of the base frequency. The weight of the partials ($w_j, j = 0,1,2,3$) were set to decrease proportionally such as $w_j = w_0 * f_p{}^j$, and the decay factor ($0 < f_p < 1$) was also a fitting parameter with a best fitting value of 0.641±0.086 (fitting error, f.e.). The best fitting bandwidth was 0.223 ± 0.088 (f.e.) Hz. The base frequency yielding the highest (positive) linear correlation with the experimental data is again the 143±15 Hz band, and the

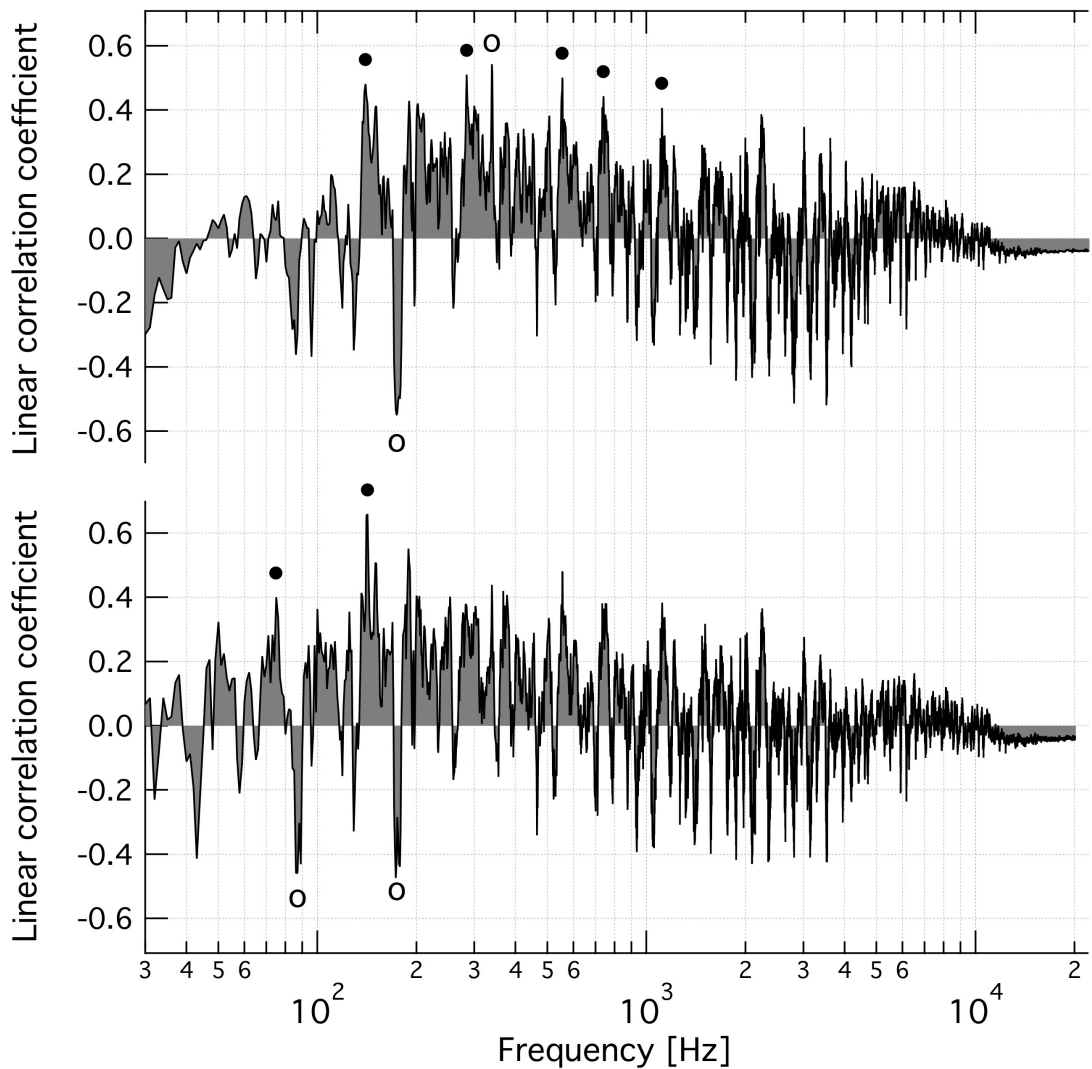

**Figure 4.** Frequency dependence of the coefficient of linear correlation between band intensity of the FFT spectra of normalised music clips and changes in specific V-ATPase activity caused by oscillating AC field of musical content, after best fitting linear scaling. Top: Circles indicate harmonically related peaks at 146±12, 285.8±8.0, 559±21, 743±31, 1127±43 Hz (solid circles) and 175.4±4.9, 338.9±4.8 Hz (open circles). The bandwidth was 1.4 Hz. Bottom: Base frequency plus the first 3 overtones. Circles indicate the harmonically related peaks at 73.4±5.2 and 143±15 Hz (solid circles) and 87.1±2.3 and 175.0±4.2 Hz (open circles). The bandwidth was 0.22 Hz. The ± frequency intervals indicate the line-width of a Gaussian function fitted to the respective peak.

largest negative correlation is again at 175.0±4.2 Hz. The first subharmonic peaks at 73.4±5.2 and 87.1±2.3 Hz are also reasonably prominent. Allowing for only even or odd overtones led to weaker correlation. If the first four overtones (as suggested by the single frequency plot, Fig. 3, top) are used and if the individual weights, relative to the base, of the overtone band are also fitting parameters then the linear correlation coefficient is 0.776, with a base frequency of 141.9 Hz and

weights ($w_j$, $j$ = 0,1,2,3) of 0.48±0.53, 1.06 ± 0.70, 1.09±0.60, and -2.4±1.3 (f.e.), for the 1st, 2nd, 3rd, and 4th overtone, respectively, and with a bandwidth of 0.210±0.050 (f.e.) Hz. The corresponding data are shown in Fig. 2 (grey circles).

Finally, an iterative search was performed without enforcing any overtone structure and with released relative weights of the frequency bands, as follows. Frequency #1 was fixed at 338.9 Hz (starting with ~146 Hz instead did not make any difference) and a plot was constructed of the linear correlation coefficient (between the $\Delta SA_i$ data and the summed intensity of two bands) with a best fitting weight of a second frequency band where the independent variable was the frequency of that band. The position of the global maximum of the plot was spotted as frequency #2. Next, frequencies #1 and #2 were fixed, and a combined intensity plot was constructed as a function of frequency #3. This procedure was repeated up to frequency #6 (there was no significant improvement above 6 bands). Then a common bandwidth was optimised and all the weights and frequency values were adjusted in a fully released fit. A linear correlation coefficient of 0.861 was obtained, the highest so far, with frequencies #1-6 of 137.12±0.22, 339.82±0.60, 1342.2±1.6, 472.3±0.25, 198.9±0.10, and 174.69±0.18 Hz and with corresponding weights of 1.03±0.38, 1, 0.57±0.17, 0.55±0.23, 0.19±0.21, and -0.64±0.18 (f.e.), and a bandwidth of 1.09±0.12 (f.e.) Hz. These scaled intensities are shown in Fig. 2 (black diamonds). Some of these frequencies are close to the harmonically related frequencies identified above. The ~199 Hz band is the least significant, with the smallest weight and a large fitting error. The ~175 Hz band has a negative weight, in agreement with the inhibitory character already identified for this band. However, new frequencies also appear in the list, suggesting that the charge movements in the rotary cycle of V-ATPase are complex and the effect of the AC field of musical origin on the specific V-ATPase activity cannot be fully described with just a few harmonically related frequency bands.

## Conclusions

It has been hypothesised long ago that membrane ATPases could be affected or even regulated by complex oscillations of the membrane potential even in living cells depending on the frequency spectrum of those oscillations[42]. An AC field generated from music represents complex oscillations with a rather wide and variable frequency spectrum. Since we have shown recently that V-ATPase shows great sensitivity to an AC field[16] of the sine waveform, it was demanding to test wether it can react differentially to AC field generated from various music of different spectral characteristics. The applied peak EF strength of 62.5 V/cm may seem to be too small for an effect on enzyme activity, however the sealed vesicle membrane amplifies the EF for membrane proteins, according to fundamental studies by Astumian and co-workers, showing that even 20 V/cm external AC field induced, e.g., active transport by Na,K-ATPase[18-20]. In our case, the amplification factor is ~1.5*300nm/5nm = 90, considering the dimensions of the membrane vesicles in our preparations[15,18]. In addition, as opposed to living cells, in our experiments all vesicles and V-ATPase molecules are exposed to the same external field, maximising the effect. Indeed, even a weaker external AC field, derived from a sine waveform, was effective on V-ATPase in the same vesicle preparations[16]. It should be noted that although V-ATPase sits in its native membrane environment, a static trans-membrane potential is not present in these vesicles. It should also be kept in mind, that although a static trans-membrane potential is present in living cells, it has been suggested that "… only an oscillating or fluctuating EF can be used by an enzyme to drive a chemical reaction away from equilibrium. In vivo, the stationary transmembrane potential of a cell must be modulated to become "locally" oscillatory if it is to derive energy and signal transduction processes."[19].

As noted already, the membrane-amplified EF depends on the angle between a line normal of the membrane surface with the EF, at the point of interest[18]. Therefore, V-ATPases, distributed homogeneously in the vesicle membrane, sense different AC field strength, with a maximum and no effect in the two poles and the equator regions, respectively, with respect to the direction defined by the electrodes. In addition, the momentary inside-outside orientation of the AC field is always opposite in the two hemispheres of a vesicle, relative to the field direction. As a consequence, the EF is in the same and opposite direction of proton transport (which is always from outside to inside the vesicle) for 50-50% of the V-ATPases, again assuming that they are homogeneously distributed in the membrane. In our previous study[16] we provided a model for how the AC field (of a sine waveform) can both inhibit and stimulate ATP hydrolysis by V-ATPase depending on the frequency of the AC field. The fact that we have observed the theoretical maximum reduction (50%) of ATPase activity for a homogenous V-ATPase distribution at very low AC frequency[16] implies that most V-ATPases are affected in our preparations.

In the present work, we found no evidence that V-ATPase would have any musical preferences with respect to genres or performers. However, the enzyme shows substantial sensitivity to certain spectral characteristics of time-dependent trans-membrane potential generated from as complex signal as music, if applied as oscillating EF, and the identified frequencies are in a perfect agreement with the observation that the musical clips that most enhance and reduce the specific activity of V-ATPase are in the musical key of D and F (or in some cases the harmonically closest 4th and or 5th intervals of them), respectively, despite the fact that the musical key and scale do not alone determine the spectrum. V-ATPase is driven by ATP hydrolysis but the AC field can affect enzymatic activity, from inhibition to stimulation. Considering the strong coupling between hydrolysis and proton pumping in V-ATPase and our previous results with sinusoidal AC[16], the proton pumping is similarly affected by musical AC as the enzymatic activity. The intensity of certain frequencies, including overtones of the estimated rate of the 60-degree rotary steps of V-ATPase, exhibit a reasonable high correlation with the effect of the AC field of musical origin on ATPase activity. This suggests that well designed specific waveforms, ideally better matching the charge motions in the catalytic rotary steps, must be more efficient in affecting V-ATPase than the pure sinusoidal waveform, which is the most widely used waveform in studies of the effect of EF on biomolecules[17,19,21]. This conclusion is in agreement with previous suggestions[17,18,20] and warrants further experiments on V-ATPase using different waveforms, preferably in real-time kinetic activity experiments. Although we found a good correlation between the experimental activity changes and band intensities of a composition of certain frequencies of the music clips, we plan further analysis of the audio clips, and also encourage other interested researchers to pursue research in this domain (we are willing to provide data and assistance), to determine whether other temporal or spectral parameters yield even better correlation with the observed "musical preferences" of V-ATPase. Finally, it should be noted that the identified prominent frequencies and musical keys are valid only under the present physical conditions under which the enzyme works. It is expected that under different conditions the control value of the specific V-ATPase activity will be different, and consequently the characteristic frequencies and musical keys will be different as well. Nevertheless, the present work demonstrates, for the first time, a musical preference of a native enzyme, a phenomenon that demands further research.

## Materials and Methods

Yeast vacuolar membrane vesicles, natively hosting high concentrations of V-ATPase, were prepared as described earlier[15]. The activity measurements with oscillating electric (AC) field applied on yeast vacuolar vesicles were also undertaken as described previously[16] except that here music was used instead of a sine waveform as the source of the AC field, and that flat electrodes were positioned, on the inner surface of a half-wide spectrometer cuvette of an optical path length of 1 cm. The assay mixture contained the activity buffer (50 mM MES/Tris, 5 mM $MgCl_2$, pH 7.0) with a starting ATP concentration of 2 mM. ATP hydrolysis was thermostated at 20±0.1 ºC for 10 minutes and then terminated. Phosphate liberated from ATP was then assayed photometrically. ATP was in excess, even at the end of the reaction period, so that the on-rate of ATP binding was not limiting the rate of ATP hydrolysis[15]. Concanamycin A (ConcA) was used at a concentration of 1 μM in order to separate the activity of V-ATPase from that of other ATPases[15]. V-ATPase constituted above 60% of the total ATPase activity in these vacuolar vesicles. The AC field was applied with flat platinum electrodes (at a distance of 4.8 mm) in a sample cell with a cross section of 5.0x10 mm, under the same conditions as for the control. Delta (± ConcA) activity values were used as a measure of ATPase activity of V-ATPase in the presence and absence of the AC field of musical origin, meaning four independent samples and a double subtraction for each experimental point in Fig. 2. The experiments were replicated 3 times on the audio clips, which were taken from published CD tracks, summarised in Table 1 and described in detail in the Supplementary Information (linked to the on-line version of the paper). If a track was longer than 5 min, the first 5 min was used unless indicated otherwise. If a track was shorter than 5 min, material was copy-pasted from the beginning to reach the 5 min clip length, without any gaps. The music clips were normalised in the digital domain to the same root-mean-square (rms) intensity of -22.53 dB. The clips were played from a Sony PCM-D50 digital recorder to a home-built analogue amplifier, which was set such that 0 dB in the digital domain corresponded to 30 V peak voltage and, consequently, -22.53 dB corresponded to 2.24 V on the electrodes. The 30 V peak voltage yielded an effective peak electric field strength of 62.5 V/cm between the electrodes. The 5 min audio clips were played twice without interruption. Metric Halo (Safety Harbor, FL 34695, USA) hardware and software were used for real-time signal analysis and generation of the audio clips containing white and pink noise.

Phosphate liberated from ATP hydrolysis in 10 min and at 20 ºC was assayed photometrically in the absence and presence of 1 μM ConcA. According to literature data[43] and our recent in-depth study on ConcA inhibition of the yeast vacuolar V-ATPase (yielding a single ConcA binding site in the same vacuolar vesicle system)[15] 1 μM ConcA is sufficient to inhibit V-ATPase but it does not affect any other, e.g. secondary transport, ATPases or P- or F-type ones (even if present in these samples, if at all). The substrate was in excess (at 2mM), and in these samples the difference in optical density is proportional to the concentration of ATP hydrolysed by V-ATPase in 10 min, that is to the specific V-ATPase activity, as we have shown earlier[15,16]. We have also shown[15] that, under the same conditions as in this work, the V-ATPase activity peaks at 2mM ATP, and that the time-dependence of ATP hydrolysis is linear (within the experimental error) over the 10 min reaction time.

Algorithms to perform numerical analysis in the frequency domain of the music audio clips were executed in Igor Pro (Wave Metrics, Lake Oswego, Oregon 97035, USA) using built-in routines, e.g. Fast Fourier Transform (FFT), statistics or curve fitting, and custom code.

**Acknowledgment**

This work was supported by the Hungarian National Research Development and Innovation Fund (K 101633, K 112716) and the GINOP-2.3.2-15-2016-00001 program (Hungary).

**Author Contributions Statement**

Pál Petrovszki carried out most of the ATPase activity measurements. Krisztina Sebők-Nagy conducted the first batch of numerical analyses in the time- and frequency domain. Tibor Páli coordinated and designed the project, carried out the final analysis of the FFT spectra, and interpreted the results.

**Additional Information**

The authors declare no competing financial or any other conflict of interests. Readers are welcome to comment on the online version of the paper. Correspondence and requests for materials should be addressed to T.P. ( tpali@brc.hu ).

**Supplementary Information**

Detailed source information of the music audio clips used in this study is given in Table S1 of the Supplementary Information document that is linked to the online version of the paper. Table S2 in the same supplementary document lists all the original experimental data.

Supplementary Information to

# The enzymatic activity of native vacuolar proton-ATPase in an oscillating electric field – demystifying an apparent effect of music on a biomolecule

Pál Petrovszki, Krisztina Sebők-Nagy, Tibor Páli*

**Table S1.** The content and source of music audio clips used in this paper, sorted from enhancement to inhibition of vacuolar proton-ATPase*

| Audio clip | Content | Source |
|---|---|---|
| 0_jarret | K. Jarret: The Köln concert, Part IIa (jazz, piano) | CD: Keith Jarret: The Köln Concert. Track #2. Performer: Keith Jarret, piano. Publisher: ECM records, München, Germany, 1975 (CD id 1064/65). (*D major, D note dominates the clip*) |
| 1_deszk | Folk songs for wine drinking, from West Hungary | CD: Mögüzenöm a rózsámnak. Track #11: Borivó nóták (Dunántúl). Performer and Publisher: Deszki Népdalkör, Deszk, Hungary, 2010 (non-commercial release). (**D major, A minor, A major**; *the pitch was not stable during the performance*) |
| 2_mozart | W.A. Mozart: Divertimento D-dur, KV251, Rondeau, Allegro Assai | CD: Music Digital Collection – 45, Mozart. Track #13. Performer: Camerata Academica Salzburg, Sandor Vegh. Publisher: GEMA, Berlin, Germany (CD id B 648). (**D major**) |
| 3_bach | J.S. Bach: Brandenburg Concertos, 5 in D-major, Allegro | CD: Zyx Classic: J.S. Bach: Brandenburgische Konzerte. Track #13: 5 in D-major, Allegro. Performer: Südwest-Studioorchester, Dir. Heribert Münchner. Publisher: Bernhard Mikulski Schallplatten-Vertriebs-GmbH, Elbtal-Dorchheim, Germany (CD id CLS 4032). (**D major**) |
| 4_deszk | Bagpipe-related folk songs from North Hungary | CD: Mögüzenöm a rózsámnak. Track #12: Felvidéki dudanóták. Performer and Publisher: Deszki Népdalkör, Deszk, Hungary, 2010 (non-commercial release). (**A major >> A minor**; *bagpipe in the key of A is ~20 cent flat*) |
| 5_legedi | Hungarian Csángó folk song: Gergely dance | CD: Legedi László István: Csobános. Track #3: Gergelytánc. Publisher: Dialekton Népzenei Kiadó, Budapest, Hungary, 2006 (CD id BS-CD 05). (*G harmonic minor or C minor, solo instrument in key of C is ~20 cent sharp, bass drum plays in C during almost the whole clip*) |
| 6_mozart | W.A. Mozart: Eine kleine Nachtmusik, 1, Allegro | CD: Music Digital Collection – 45, Mozart. Track #1. Serenade No. 13 G-dur KV 525. Performer: Franz-Liszt-Kammerorchester, János Rolla. Publisher: Music GEMA, Berlin, Germany (CD id B 648). (**G major**) |
| 7_berry | C. Berry: Back in the USA | CD: Chuck Berry. Track #10 (the first 2.5 minutes is repeated). Performer: Chuck Berry. Publisher: Signal, Israel, (CD id 50650). (*Eb major, but in 430Hz tuning*) |
| 8_boneym | Boney M: Megamix | CD: The greatest hits, Boney M. Track #14: Boney M Megamix: Rivers of Babylon, Sunny, Daddy Cool, Rasputin (the first 2.5 minutes is repeated). Performer: Boney M. Publisher: Telstar Records, Plc., 1992 (CD id TCD 2656). (*C major, C minor, E minor*) |
| 9_bach | J.S. Bach: Toccata and Fugue in d-minor | CD: Die Grossen Meister der Klassischen Musik. Track #8. Performer: Hannes Kästner. Publisher: Delta Music GmbH, Königsdorf, Germany, 1993 (CD id. 11501). (**D minor**) |
21

| | | |
|---|---|---|
| 10_fchoir | Female choir: rearranged Hungarian folk songs | Concert CD. Track #16: Karai J.: Estéli nótázás. Performer and Publisher: Szeghy Endre Pedagógus Női Kar, Szeged, Hungary, 2012 (non-commercial release). (**G minor >> D minor > G mixolydian**) |
| 11_liszt | F. Liszt: Buch der Lieder für Piano allein, No. 1 | CD: Liszt Ferenc, Piano Concerto, Opus Postumum. Track #6. Performer: Jenő Jandó, Hungarian State Orchestra, Lamberto Gardelli. Publisher: Hungaroton, Budapest, Hungary, 1991 (CD id HCD 31396). (*E major >> G major > F-sharp major*) |
| 12_liszt | F. Liszt: Concerto for Piano & Orchestra in E flat major, 3, Andante | CD: Liszt Ferenc, Piano Concerto, Opus Postumum. Track #3. Performer: Jenő Jandó, Hungarian State Orchestra, Lamberto Gardelli. Publisher: Hungaroton, Budapest, Hungary, 1991 (CD id HCD 31396). (**E flat major**) |
| 13_wnoise | White noise (not music) | Generated with software. |
| 14_abba | ABBA: Waterloo | CD: The ABBA Remasters, ABBA. Track #12 (the first 2.5 minutes is repeated). Performer: ABBA. Publisher: Polar Music International B.V., PolyGram, 1997 (CD id 533983-2). (*D major*) |
| 15_jarre | Jean-Michel Jarre: Arpegiator (synthesiser music) | CD: Jean-Michel Jarre, Musik aus Zeit und Raum. Track #5. Performer: Jean-Michel Jarre. Publisher: PolyGram, Hanover, Germany (CD id 815686-2). (*C minor*) |
| 16_pnoise | Pink noise (not music) | Generated with software. |
| 17_bach | J.S. Bach: Brandenburg Concertos, 2, in F-major, Allegro | CD: Zyx Classic, J.S. Bach: Brandenburgische Konzerte. Track #5. Performer: Südwest-Studioorchester, Dir. Heribert Münchner. Publisher: Bernhard Mikulski Schallplatten-Vertriebs-GmbH, Elbtal-Dorcheim, Germany (CD id CLS 4032). (**F major**) |
| 18_deszk | Wedding folk music from South Hungary | CD: Mögüzenöm a rózsámnak. Track #4: Délvidéki lakodalmas (Alföld). Performer and Publisher: Deszki Népdalkör, Deszk, Hungary, 2010 (non-commercial release). (**A minor > C major, A major**) |
| 19_vivaldi | A. Vivaldi: The Four Seasons, Autumn, 1, Allegro | CD: Music Digital Collection – 15, Vivaldi. Track #7: Concerto No. 3 F-major, RV 293, "Autumn", 1, Allegro. Performer: Budapest Strings, Károly Botvay. Publisher: GEMA, Berlin, Germany (CD id B 518). (**F major**) |

*The most frequented key(s) and scale(s) of the clips are given (referencing to 440Hz tuning, except when indicated otherwise) in boldface for those tracks for which they were indicated on the CD or provided by the artists, and in italics for those determined by the authors (with the help of professional musicians). Only up to the three most dominant keys and scales are given for each clip. The ">" symbol indicates the relative abundance of the given keys and scales in the clip.

**Table S2.** Optical densities (*OD*) and specific ATPase activities* (*SA* in nmole ATP min$^{-1}$ mg$^{-1}$) in vacuolar vesicle dispersions (all experiments)

| Audio clip | +AC, -ConcA | | +AC, +ConcA | | -AC, -ConcA | | -AC, +ConcA | |
|---|---|---|---|---|---|---|---|---|
| | *OD* | *SA* | *OD* | *SA* | *OD*** | *SA*** | *OD*** | *SA*** |
| 0_jarret | 1.219 | 74.388 | 0.494 | 20.916 | 1.157 | 69.815 | 0.57 | 26.521 |
| 0_jarret | 1.213 | 73.945 | 0.513 | 22.317 | 1.157 | 69.815 | 0.57 | 26.521 |
| 0_jarret | 1.194 | 72.544 | 0.551 | 25.120 | 1.157 | 69.815 | 0.57 | 26.521 |
| 1_deszk | 1.07 | 63.399 | 0.414 | 15.016 | 1.057 | 62.440 | 0.458 | 18.261 |
| 1_deszk | 1.084 | 64.431 | 0.41 | 14.721 | 1.057 | 62.440 | 0.458 | 18.261 |
| 1_deszk | 1.091 | 64.947 | 0.405 | 14.352 | 1.057 | 62.440 | 0.458 | 18.261 |
| 2_mozart | 1.15 | 69.299 | 0.498 | 21.211 | 1.157 | 69.815 | 0.57 | 26.521 |
| 2_mozart | 1.182 | 71.659 | 0.511 | 22.170 | 1.157 | 69.815 | 0.57 | 26.521 |
| 2_mozart | 1.157 | 69.815 | 0.524 | 23.129 | 1.157 | 69.815 | 0.57 | 26.521 |
| 3_bach | 1.046 | 61.628 | 0.403 | 14.204 | 1.057 | 62.440 | 0.458 | 18.261 |
| 3_bach | 1.037 | 60.965 | 0.398 | 13.836 | 1.057 | 62.440 | 0.458 | 18.261 |
| 3_bach | 1.043 | 61.407 | 0.424 | 15.753 | 1.057 | 62.440 | 0.548 | 24.899 |
| 4_deszk | 1.078 | 63.989 | 0.418 | 15.311 | 1.057 | 62.440 | 0.458 | 18.261 |
| 4_deszk | 1.104 | 65.906 | 0.432 | 16.343 | 1.057 | 62.440 | 0.458 | 18.261 |
| 4_deszk | 1.067 | 63.177 | 0.415 | 15.089 | 1.057 | 62.440 | 0.458 | 18.261 |
| 5_legedi | 1.183 | 71.733 | 0.536 | 24.014 | 1.157 | 69.815 | 0.57 | 26.521 |
| 5_legedi | 1.147 | 69.078 | 0.51 | 22.096 | 1.157 | 69.815 | 0.57 | 26.521 |
| 5_legedi | 1.168 | 70.627 | 0.515 | 22.465 | 1.157 | 69.815 | 0.57 | 26.521 |
| 6_mozart | 1.179 | 71.438 | 0.517 | 22.612 | 1.157 | 69.815 | 0.57 | 26.521 |
| 6_mozart | 1.168 | 70.627 | 0.568 | 26.374 | 1.157 | 69.815 | 0.57 | 26.521 |
| 6_mozart | 1.162 | 70.184 | 0.487 | 20.400 | 1.157 | 69.815 | 0.57 | 26.521 |
| 7_berry | 1.062 | 62.809 | 0.418 | 15.311 | 1.057 | 62.440 | 0.458 | 18.261 |
| 7_berry | 1.064 | 62.956 | 0.435 | 16.565 | 1.057 | 62.440 | 0.458 | 18.261 |
| 7_berry | 1.07 | 63.399 | 0.403 | 14.204 | 1.057 | 62.440 | 0.458 | 18.261 |
| 8_boneym | 1.031 | 60.522 | 0.423 | 15.679 | 1.057 | 62.440 | 0.458 | 18.261 |
| 8_boneym | 1.054 | 62.219 | 0.415 | 15.089 | 1.057 | 62.440 | 0.458 | 18.261 |
| 8_boneym | 1.048 | 61.776 | 0.398 | 13.836 | 1.057 | 62.440 | 0.458 | 18.261 |
| 9_bach | 1.04 | 61.186 | 0.403 | 14.204 | 1.057 | 62.440 | 0.458 | 18.261 |
| 9_bach | 1.022 | 59.858 | 0.43 | 16.196 | 1.057 | 62.440 | 0.458 | 18.261 |
| 9_bach | 1.06 | 62.661 | 0.435 | 16.565 | 1.057 | 62.440 | 0.458 | 18.261 |
| 10_fchoir | 1.175 | 71.143 | 0.573 | 26.743 | 1.157 | 69.815 | 0.57 | 26.521 |
| 10_fchoir | 1.209 | 73.650 | 0.631 | 31.020 | 1.157 | 69.815 | 0.57 | 26.521 |
| 10_fchoir | 1.171 | 70.848 | 0.569 | 26.448 | 1.157 | 69.815 | 0.57 | 26.521 |
| 11_liszt | 1.166 | 70.479 | 0.516 | 22.539 | 1.157 | 69.815 | 0.57 | 26.521 |
| 11_liszt | 1.137 | 68.340 | 0.607 | 29.250 | 1.157 | 69.815 | 0.57 | 26.521 |
| 11_liszt | 1.148 | 69.151 | 0.569 | 26.448 | 1.157 | 69.815 | 0.57 | 26.521 |
| 12_liszt | 1.164 | 70.332 | 0.604 | 29.029 | 1.157 | 69.815 | 0.57 | 26.521 |
| 12_liszt | 1.145 | 68.930 | 0.578 | 27.111 | 1.157 | 69.815 | 0.57 | 26.521 |
| 12_liszt | 1.152 | 69.446 | 0.541 | 24.382 | 1.157 | 69.815 | 0.57 | 26.521 |
| 13_wnoise | 1.09 | 64.874 | 0.446 | 17.376 | 1.024 | 60.006 | 0.371 | 11.844 |

| 13_wnoise | 1.078 | 63.989 | 0.404 | 14.278 | 1.024 | 60.006 | 0.371 | 11.844 |
| 13_wnoise | 1.00 | 58.236 | 0.389 | 13.172 | 1.024 | 60.006 | 0.371 | 11.844 |
| 14_abba | 0.961 | 55.359 | 0.342 | 9.705 | 1.024 | 60.006 | 0.371 | 11.844 |
| 14_abba | 1.018 | 59.563 | 0.376 | 12.213 | 1.024 | 60.006 | 0.371 | 11.844 |
| 14_abba | 0.962 | 55.433 | 0.369 | 11.697 | 1.024 | 60.006 | 0.371 | 11.844 |
| 15_jarre | 1.16 | 70.037 | 0.562 | 25.931 | 1.157 | 69.815 | 0.57 | 26.521 |
| 15_jarre | 1.135 | 68.193 | 0.597 | 28.513 | 1.157 | 69.815 | 0.57 | 26.521 |
| 15_jarre | 1.099 | 65.537 | 0.605 | 29.103 | 1.157 | 69.815 | 0.57 | 26.521 |
| 16_pnoise | 0.946 | 54.253 | 0.473 | 19.367 | 1.024 | 60.006 | 0.371 | 11.844 |
| 16_pnoise | 0.981 | 56.834 | 0.365 | 11.402 | 1.024 | 60.006 | 0.371 | 11.844 |
| 16_pnoise | 0.93 | 53.073 | 0.377 | 12.287 | 1.024 | 60.006 | 0.371 | 11.844 |
| 17_bach | 0.922 | 52.483 | 0.467 | 18.925 | 1.024 | 60.006 | 0.371 | 11.844 |
| 17_bach | 1.013 | 59.195 | 0.461 | 18.482 | 1.024 | 60.006 | 0.371 | 11.844 |
| 17_bach | 0.984 | 57.056 | 0.42 | 15.458 | 1.024 | 60.006 | 0.371 | 11.844 |
| 18_deszk | 1.127 | 67.603 | 0.668 | 33.749 | 1.157 | 69.815 | 0.57 | 26.521 |
| 18_deszk | 1.162 | 70.184 | 0.685 | 35.003 | 1.157 | 69.815 | 0.57 | 26.521 |
| 18_deszk | 1.152 | 69.446 | 0.738 | 38.912 | 1.157 | 69.815 | 0.57 | 26.521 |
| 19_vivaldi | 0.974 | 56.318 | 0.477 | 19.662 | 1.024 | 60.006 | 0.371 | 11.844 |
| 19_vivaldi | 0.996 | 57.941 | 0.537 | 24.087 | 1.024 | 60.006 | 0.371 | 11.844 |
| 19_vivaldi | 0.964 | 55.581 | 0.535 | 23.940 | 1.024 | 60.006 | 0.371 | 11.844 |

*Calculated using the equation $SA = (OD - 0.21041)/(0.0045195 * 0.3 * 10)$ nmole ATP min$^{-1}$ mg$^{-1}$ based on the total protein content (0.3 mg) per sample, 10 min reaction time and the inorganic phosphate calibration in the same system published earlier[15].

**Identical numbers for the control (no AC filed) samples are those of the corresponding isolation batches (several samples were made from each batch).

+ + +